\documentclass{emulateapj}

\DeclareRobustCommand{\ion}[2]{%
\relax\ifmmode
\ifx\testbx\f@series
{\mathbf{#1\,\mathsc{#2}}}\else
{\mathrm{#1\,\mathsc{#2}}}\fi
\else\textup{#1\,{\mdseries\textsc{#2}}}%
\fi}

\slugcomment{To appear in ApJ.}

\shorttitle{The star formation history of NGC 4435} \shortauthors{Panuzzo et al.}

\begin{document}

\title{The Star Formation History of the Virgo early-type galaxy NGC~4435: the
{\it Spitzer} Mid Infrared  view.}

\author{P. Panuzzo\altaffilmark{1}, O. Vega\altaffilmark{2},
A. Bressan\altaffilmark{1,2,4}, L. Buson\altaffilmark{1}, M. Clemens\altaffilmark{1}, 
R. Rampazzo\altaffilmark{1},\\
L. Silva\altaffilmark{3}, J. R. Vald\'es\altaffilmark{2}, G. L. Granato\altaffilmark{1,4},
L. Danese\altaffilmark{4}}

\altaffiltext{1}{INAF Osservatorio Astronomico di Padova, vicolo dell'Osservatorio 5, 35122 Padova, 
Italy; pasquale.panuzzo@oapd.inaf.it, alessandro.bressan@oapd.inaf.it, lucio.buson@oapd.inaf.it, 
marcel.clemens@oapd.inaf.it, gianluigi.granato@oapd.inaf.it, roberto.rampazzo@oapd.inaf.it}

\altaffiltext{2}{Instituto Nacional de Astrof\'{\i}sica, Optica y Electr\'onica,
Apdos. Postales 51 y 216, C.P. 72000 Puebla, Pue., M\'exico; ovega@inaoep.mx, jvaldes@inaoep.mx}

\altaffiltext{3}{INAF Osservatorio Astronomico di Trieste, Via Tiepolo 11,
I-34131 Trieste, Italy; silva@oats.inaf.it}

\altaffiltext{4}{Scuola Internazionale Superiore de Studi Avanzati (SISSA),
via Beirut 4, 34014, Trieste, Italy; danese@sissa.it}


\begin{abstract}
We present a population synthesis study
of NGC~4435, an early-type Virgo galaxy interacting with NGC~4438.
We combine new spectroscopic observations obtained with the
\emph{Spitzer} Space Telescope IRS instrument with 
IRAC archival data and broad band data from the literature.
The IRS spectrum shows prominent PAH features, low ionization emission
lines and H$_2$ rotational lines arising
from the dusty circumnuclear disk characterizing this
galaxy. The central SED, from X-ray to radio, is well fitted by a model
of an exponential burst superimposed on an old simple stellar population.
From the lack of high excitation
nebular lines, the [Ne{\sc\,iii}]15.5/[Ne{\sc\,ii}]12.8 ratio,
the temperature of molecular hydrogen, and the fit to the full X-ray to radio
SED we argue that the present activity of the galaxy is driven
by star formation alone. The AGN contribution to the ionizing
flux is constrained to be less than 2\%.
The age of the burst is found to be around 190 Myr
and it is fully consistent with the notion
that the star formation process has been
triggered by the interaction with NGC~4438.
The mass involved in the rejuvenation
episode turns out to be less than 
1.5\% of the stellar galaxy mass sampled in a 5\arcsec\ central aperture.
This is enough to render NGC~4435
closely similar to a typical interacting early-type galaxy with
inverted Ca{\sc\,ii}[H+K] lines
that will later turn into a typical {\sl cluster} E+A galaxy
and enforces the notion that these objects are the result of a recent 
rejuvenation episode rather than a genuine delayed formation.
\end{abstract}



\section{Introduction}
\label{sec:intro}

NGC~4435  (VCC1030), an early-type  SB0$_1$(7)  galaxy \citep{sand81}
member of the Virgo Cluster, is of great interest because it is a 
nearby example of galaxy interactions in the cluster environment.
NGC~4435 is experiencing an off-center encounter with NGC~4438,
a bulge dominated late-type galaxy lying at a projected separation of
4.3\arcmin\  \citep[i.e. 20 kpc, adopting a Virgo distance of 16.1 Mpc from][]{kels00}.
Evidence of the strong on-going interaction comes from the well developed stellar
tidal tail ($\approx$30 kpc) of NGC~4438 but also from the structure of 
the inter stellar medium (ISM) of both galaxies.  H$\alpha$+[\ion{N}{ii}]
images \citep{kenn95} reveal several ionized  gas filaments starting from the disk of  NGC~4438
and extending out of the disk up to 10 kpc. The study of the distribution of the cold gas
shows that  CO and \ion{H}{i} are displaced west of the center of NGC~4438, although
CO is present also in the center of NGC~4435. \citet{voll05} proposed a model
of the encounter based on the combined picture coming from the ISM  observations and
simulations. These latter authors indicate that the evolution of NGC~4438 has been influenced by
its passage though the intra-cluster medium (ICM), where ram-pressure stripped the \ion{H}{i} in about
100 Myr. During this period, NGC~4435 passed through the disk of NGC~4438 at high velocity
($\approx$ 800 km~s$^{-1}$), but the ISM-ISM interaction of the two galaxies had a less
important effect on the final distribution of the gas than the cluster ram-pressure stripping
exerted by the ICM.

If the above is the general picture of the NGC~4435/38 encounter, its effects on the
evolution of NGC~4435 are less clear. 
In low density environments, \citet{domi03}
showed that early-type galaxies show little or no IR or optical signatures
of star formation when physically paired to a gas rich companion.
However, some of the early-type members do show significant levels of IR
and H$\alpha$ emission that indicate both thermal (star formation) and non thermal
(AGN) activity. These pairs are among the best candidates for direct
interaction fueling of both starburst and active galactic nuclei.

NGC~4435 currently has an ISM. HST images reveal the presence of a
circumnuclear disk of 4\arcsec\ radius \citep[360 pc, inclination 44 deg;][]{hosa02,cape05}.
Moreover, nebular emission lines were detected by several
authors \citep{kenn95,hofi95,hofi97}; in particular \citet{hofi97},
using the classical diagnostic diagrams, 
classified this galaxy as T2/H:, i.e. a transition object between LINER 2 and 
\ion{H}{ii}.

\citet{cocc06} studied the ionized gas kinematics using HST-STIS
in the central 2\arcsec\ looking for a super massive
black hole (SMBH). The rotation curve is very symmetric, indicating a regular
rotation. Their analysis indicates that the mass of the SMBH (upper
limit $7.5\times 10^6~ M_\odot$) is lower than that expected from the $M_{\rm BH}$-$\sigma$
and $M_{\rm BH}$-$L_{\rm bulge}$ relations. Coccato et al. suggest that the SMBH in
NGC~4435 could be of the ``laggard'' type as discussed in \citet{vitt05},
although these seem to lurk among galaxies that spend most of their
lifetime in the field, at least apparently at odds with NGC~4435's cluster
location.

In this paper we study the \emph{Spitzer} IRS
MIR spectra of NGC~4435 presented in \citet{bres06} and 
perform a comprehensive  analysis of the entire SED of the
galaxy. In particular, the study of the MIR region, rich in
spectral features (emission lines, PAH, dust emission etc.)
provides an independent view of the nature of the nuclear activity
of this galaxy \citep{genz98,lutz98,rous01}.
Combined with the analysis of the panchromatic SED, it allows the characterization of the 
star formation activity and the estimate of the mass of stars
formed during this interaction, shedding light on 
the rejuvenation mechanism in interacting early-type galaxies.

The paper is organized as follows. In \S~\ref{sec:observ} we provide details of the observations
and data reduction methods. In \S~\ref{sec:miranalysis} we describe the results of the study
of the spectrum in the MIR region, while in \S~\ref{sec:sedglobal} we extend the modeling to the
entire galaxy SED. In \S~\ref{sec:discussion} we discuss the result of our analysis and compare them
with other studies. Finally in \S~\ref{sec:conclu} we summarize the conclusions.


\section{Observations and data reduction}
\label{sec:observ}

\subsection{IRS spectra}
\label{sec:irs_redu}

{\em Spitzer} IRS spectral observations of NGC~4435 were obtained during
the first {\em Spitzer} General Observer Cycle on 2005 June 1
as part of a program \citep[ID 3419,][]{bres06}  to study
early-type galaxies that belong to the colour -- magnitude relation of
the Virgo cluster \citep{bowe92}. 

The observations were performed
in Standard Staring mode with low resolution ($R\sim$
64--128) modules SL1 (7.5--15.3$\mu$m), SL2
(5--7.6$\mu$m) and LL2 (14.1--21.3$\mu$m), 
with 6 exposures of 60 seconds for each SL module and 10 exposures
of 120 seconds for the LL2 module.

Since the adopted IRS pipeline (version S12) is
specifically designed for point source flux extraction, we have
devised a new procedure to flux calibrate the spectra that
exploits the large degree of symmetry that characterizes the light
distribution in early-type galaxies.
The reduction procedure is described in detail in
\citet{bres06} thus here we only summarize the main steps.

For SL observations, the sky background was removed by subtracting
observations taken in different orders, but at the same node
position. LL segments were sky-subtracted by differencing the two
nod positions.

We obtained new e$^-$/s to Jy flux conversions by applying a correction
for aperture losses (ALCF) and a correction for slit losses (SLCF)
to the flux conversion tables provided by the
Spitzer Science Center \citep[e.g.][]{kenn03}.
By applying the ALCF and SLCF corrections, we obtained the flux {\sl
received} by the slit.

For each galaxy of the program, we simulated the corresponding
observed linear profile along the slits  by
convolving
a wavelength dependent bidimensional intrinsic surface brightness profile
with the instrumental point spread function (PSF).
The adopted profile is a two dimensional
modified King law \citep{elso87}.
By fitting the observed profiles with the simulated ones, we can reconstruct
the intrinsic profiles and the corresponding intrinsic SED.
This procedure has also the advantage  to recognize
whether a particular feature is spatially extended or not.

\begin{figure}
\begin{center}
\plotone{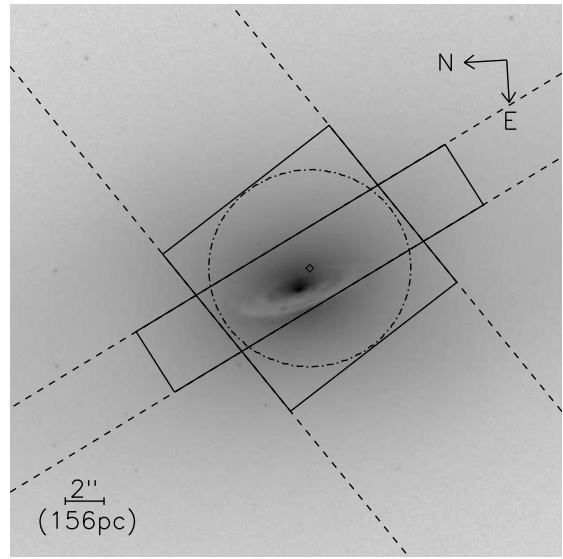}
\caption{HST WFPC2/F814W image of the NGC~4435 nucleus \citep{cocc06} with
IRS SL and LL slits (dashed lines) during our observations.
The rectangles show the aperture used to extract the spectrum. The dot-dashed circle
shows the 5\arcsec\ circular aperture used to define the SED of the central
region from broad band data (see Sec. \ref{sec:sedglobal}).}
\label{fig:hst}
\end{center}
\end{figure}

Finally the spectrum was extracted in a fixed width (18\arcsec\ for SL and
10.2\arcsec\ for LL) around the maximum intensity.
In order to account for the different
extraction area, the LL spectrum has
been rescaled to match that of SL (3.6\arcsec$\times$18\arcsec).
In Figure \ref{fig:hst} we overplot the extraction area and the IRS
slit positions on an HST WFPC2/F814W image of the NGC~4435 nucleus by
\citet{cocc06}.

The uncertainty on the flux was evaluated by
considering two sources of noise: the instrumental plus background noise
and the poissonian noise of the source. The former was evaluated
by measuring the variance of pixel values in background-subtracted coadded
images far from the source.
The poissonian noise of the sources was estimated as the square root of the
ratio between the variance of the number of e$^-$ extracted per pixel in 
each exposure, and the number of the exposures.
The total noise was obtained by summing the two sources in quadrature 
and by multiplying by the square root of the
extraction width in pixels.
We notice that the overall absolute photometric uncertainty
of IRS is 10\%, while the slope deviation within
a single segment (affecting all spectra in the same way)
is less  than 3\% (see the Spitzer Observer Manual).

\subsection{IRAC and MIPS images}
\label{irac}
IRAC and MIPS images of NGC~4435 were retrieved from the SSC archive. They
were taken as part of the program ID 3649. Fluxes at 3.6, 4.5, 5.8
and 8~$\mu$m were measured from the IRAC images as output
from the ``post-BCD'' pipeline. The proximity of NGC~4435 to NGC~4438
results in a weak contaminating flux component. This was treated
by estimating the median background flux from the image containing
NGC~4435 so that the subtracted background included the
contaminating light. Fluxes were extracted in circular apertures
centered on the brightest pixel at each wavelength. Given that the
calibration of extended sources for IRAC is still somewhat uncertain the
fluxes so extracted have errors of $\sim 10\%$ \citep{dale05}.
Similarly, we measured fluxes at 24, 70 and 160~$\mu$m from 
MIPS mosaics as output from the ``post-BCD'' pipeline. 
Fluxes were extracted in circular apertures centered on the brightest 
pixel at each wavelength, subtracting the median background flux
measured far from the source.
Uncertainties for MIPS fluxes are 
quoted to be $\sim 20\%$ \citep[][see also the Spitzer Observer Manual]{dale05}.
It is worth noting that in all MIPS images the 
galaxy appares almost as a point source; in particular at 24~$\mu$m where the 
angular resolution is higest the galaxy image is only slightly larger than the
instrumental PSF (6\arcsec).

\section{Analysis of the mid-IR spectrum}
\label{sec:miranalysis}

\begin{figure*}
\plottwo{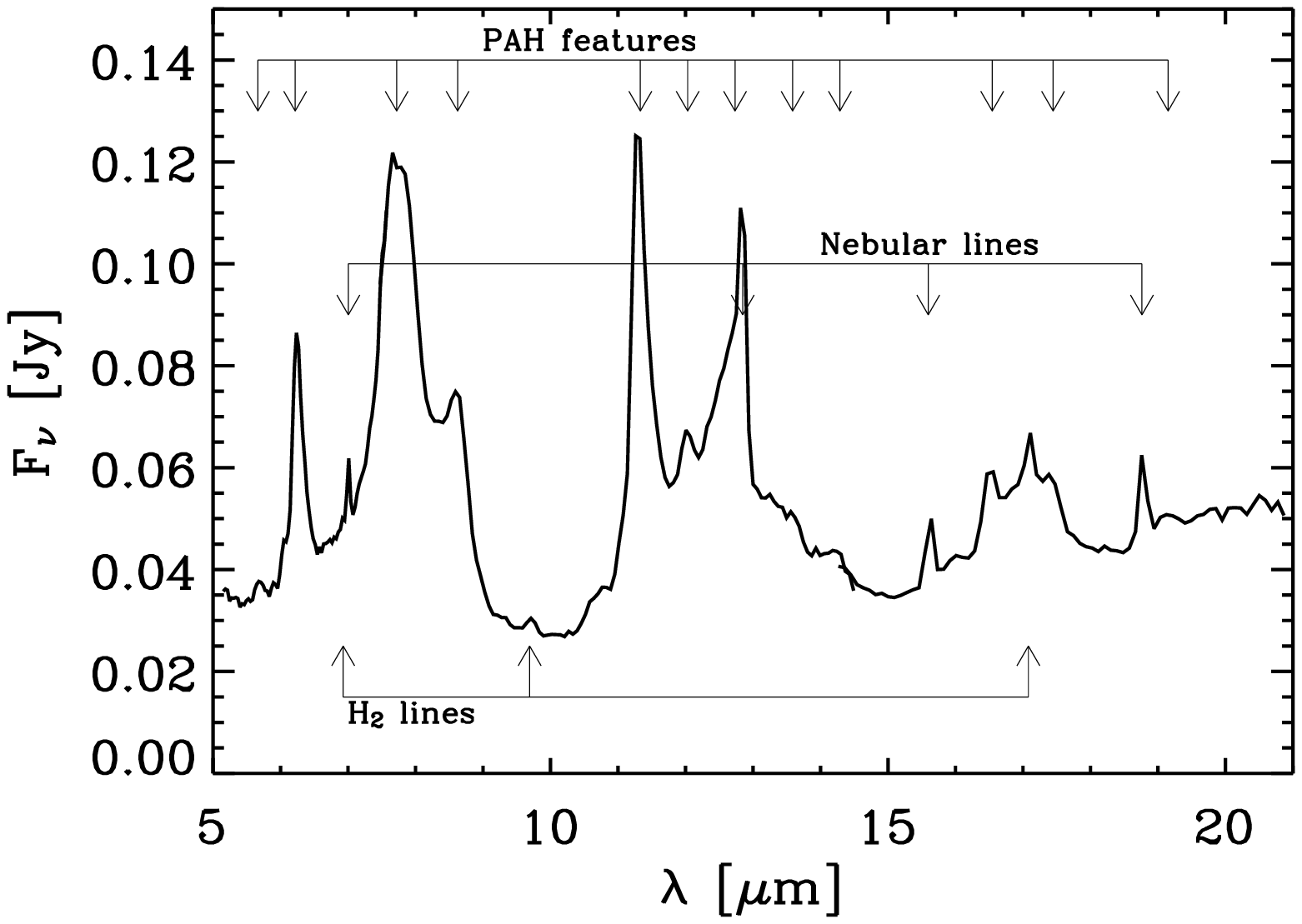}{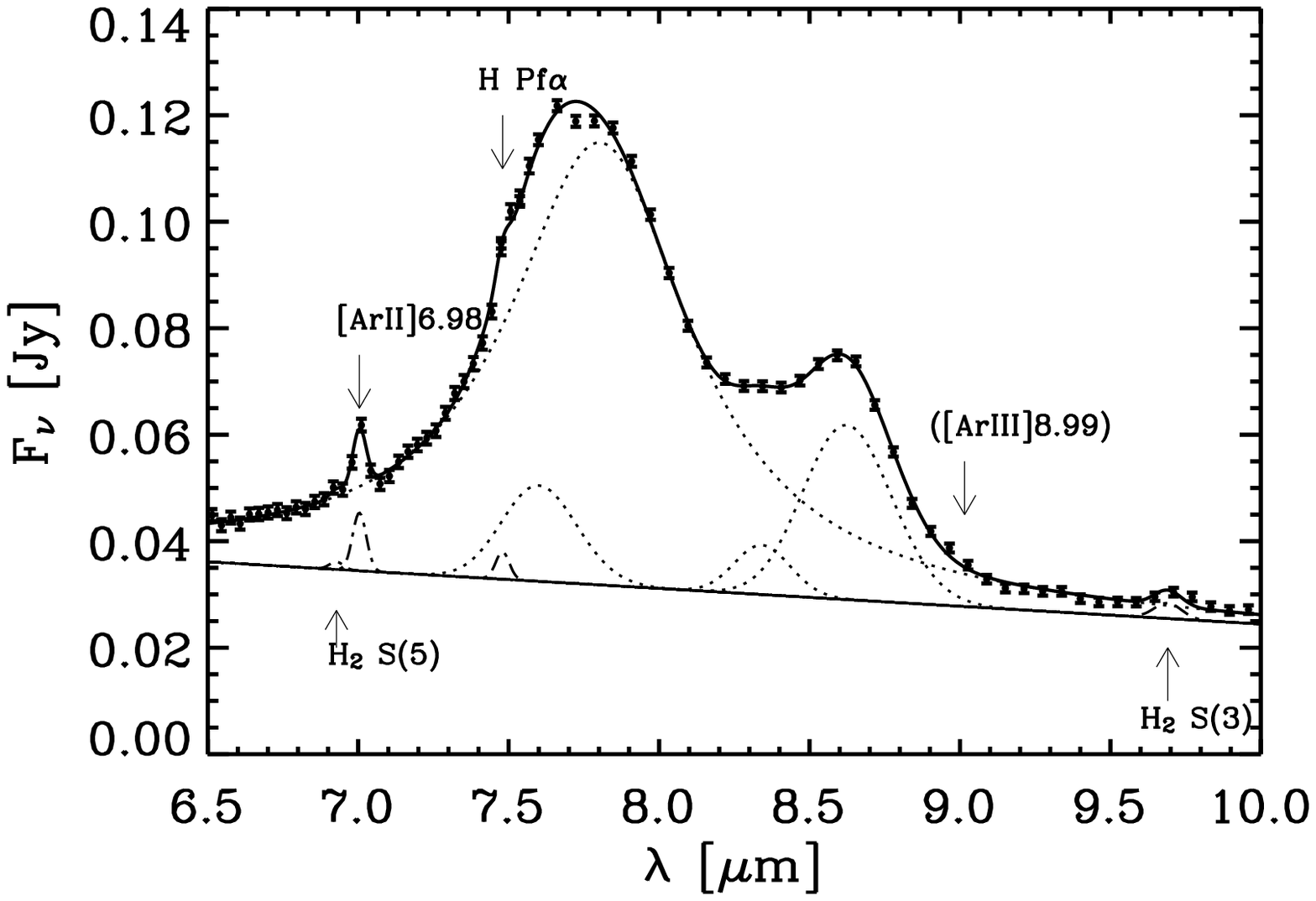}\\
\plottwo{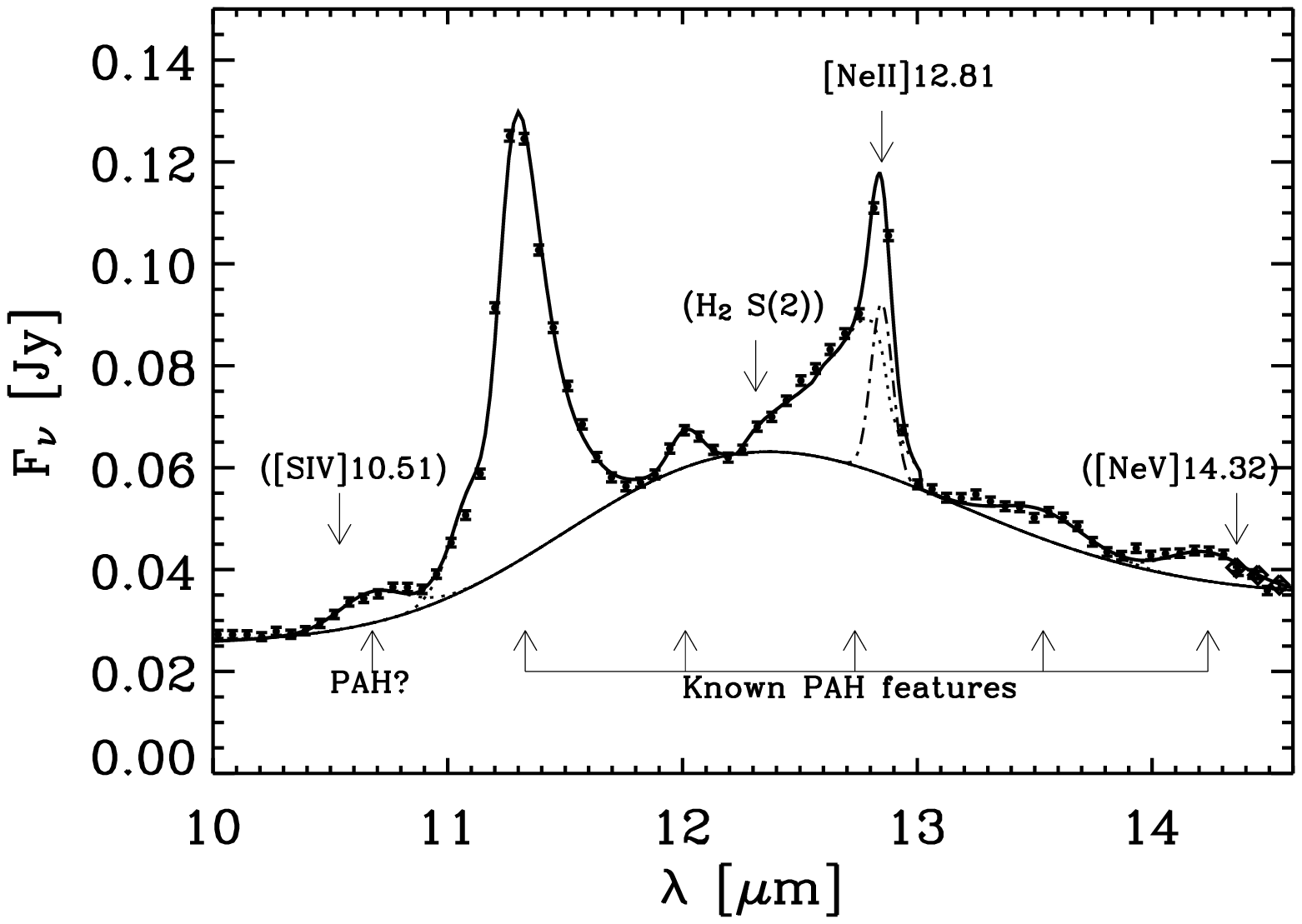}{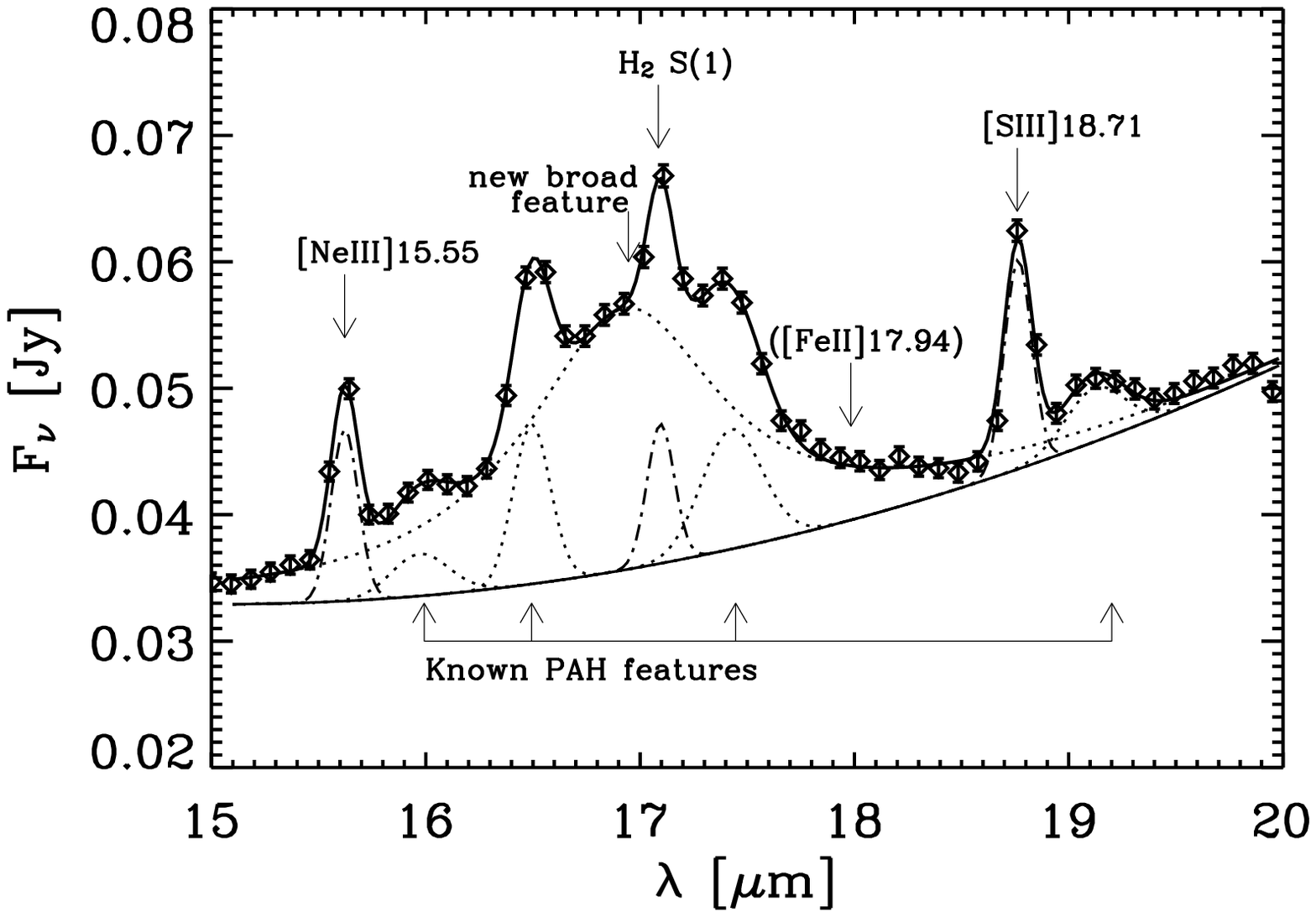}
\figcaption{\emph{Spitzer} IRS MIR spectrum of NGC~4435. {\em
Top-left:} entire observed spectrum; arrows indicate the positions
of PAH features, nebular emission lines and H$_2$ rotational lines.
{\em Top-right:} the 6.5--10$\mu$m spectral region split in the
different components.
The error bars  correspond to 1 $\sigma$ noise fluctuations.
The lower smooth solid
line corresponds to the continuum component; the dotted line to the
continuum plus PAH features, the dot-dashed lines show the continuum plus
emission lines. Finally the upper solid line represents the sum of all
components.
Lines for which we could provide only upper
limits are included in parenthesis. {\em Bottom-left:} similar to
top-right panel but for the 10-14.6$\mu$m spectral region. {\em
Bottom-right:} similar to top-right panel but for the 15-20$\mu$m
spectral region.} \label{fig:spectrum_tot}
\end{figure*}

The final flux calibrated IRS spectrum of NGC~4435 is shown in
the  top-left panel of Figure \ref{fig:spectrum_tot}.
The spectrum is dominated by numerous
PAH features and shows several
nebular emission lines, [\ion{Ar}{ii}]7$\mu$m, [\ion{Ne}{ii}]12.8$\mu$m,
[\ion{Ne}{iii}]15.5$\mu$m and [\ion{S}{iii}]18.7$\mu$m, and molecular hydrogen
rotational lines, H$_2$ S(3) 9.66$\mu$m, H$_2$ S(1) 17.04$\mu$m.

In order to measure the fluxes of emission lines and to identify the PAH
features, we fitted the spectrum via decomposition into various components:
PAH emission features, emission lines and underlying continuum.

For all the PAH features but those at 11.2 and 12.5 $\mu$m, we
assumed a Drude profile. The features at 11.2 and 12.5 $\mu$m
are very asymmetric thus we have used the
profiles observed by \citet{hony01}, convolved to the SL1
resolution. It is worth noticing that the emission profiles derived
by \citet{hony01} do not account for the blending of the wings
of the PAH features. Consequently our underlying continuum shows a
broad feature. However, since the main purpose of this section is to
estimate the emission line intensities, this does
not affect our conclusions.

For the emission lines we adopted a
Gaussian profile, with a FWHM given by the resolving power of the
corresponding IRS module (see the \emph{Spitzer} Observer's Manual,
v. 6.0, sec. 7.1.6).
The resulting fits of the three most interesting spectral regions are shown
in Figure \ref{fig:spectrum_tot}.


\subsection{PAH features}
\label{sec:PAH_features}

In the IRS spectrum of NGC~4435 we can identify many PAH emission
features already known from ISO and pre-ISO studies: at 6.2, 7.7,
8.2, 8.6, 11.2, 11.9, 12.7, 13.5, 14.5, 15.9, 16.4~$\mu$m 
\citep[e.g.][]{lidr01,stur00}. 
Besides those features, we
also see a number of new components that have been revealed by
recent \emph{Spitzer} observations. These are the feature at
10.7~$\mu$m and, most notably, the complex around 17~$\mu$m. The
nature of the former is unknown but there is similar evidence in
M82 \citep{fors04}, NGC~7714 \citep{bran04} and
NGC~7331 \citep{smit04} spectra.
The complex around 17~$\mu$m (Figure
\ref{fig:spectrum_tot}, bottom-right panel) is similar to that
observed in the spiral galaxies NGC~7331 and
in the extended emission of M82
\citep{enge06}. This complex is superimposed on H$_2$
S(1) emission but in our case, since the H$_2$ S(1) emission is
fainter than that seen in NGC~7331 and M82, we can more
accurately extract the underlying emission components. In fact,
while  \citet{smit04} claimed the presence of a new broad
emission feature at 17.1~$\mu$m with a FWHM of 0.96~$\mu$m, in our
spectrum this feature appears to be a blend of two components; a narrow
component centered at 17.4~$\mu$m with a FWHM of 0.33~$\mu$m and
a broad component centered at 16.9~$\mu$m with a FWHM of 1.1~$\mu$m.
Notice that the component centered at 17.4~$\mu$m was observed by
\citet{vank00} in various galactic star formation regions.
Thus our observations indicate that
the new PAH emission feature is centered at 16.9~$\mu$m. In the
same spectral region we also find an emission feature at 19.07~$\mu$m
that was observed in the NCG~7023 reflection nebula by \citet{wern04}.


\subsection{Emission lines}
In the MIR spectrum of NGC~4435 we
detect only low ionization emission lines. The corresponding
intensities are listed in table \ref{tab:emlines}. The reported
uncertainties in the line fluxes are 1$\sigma$ as derived from
the line fitting.
For some interesting higher excitation emission lines we can only provide
$3\sigma$ upper limits. In the following we discuss the main
implications of our detections.

\begin{deluxetable}{lccc}
\tablewidth{0pt}
\tablecaption{Emission line intensities.\label{tab:emlines}}
\tablehead{
\colhead{Transition/Ion}& 
\colhead{$\lambda_{\rm rest}$}& 
\colhead{Flux}\\
\colhead{}& 
\colhead{($\mu$m)} & 
\colhead{($10^{-17}$ W m$^{-2}$)}
}
\startdata
H$_2$ 0--0 S(5)   &\phn6.9095 & 0.505 $\pm$ 0.411\\
$[$\ion{Ar}{ii}]  &\phn6.9853 & 3.787 $\pm$ 0.378\\
H Pf$\alpha$      &\phn7.4599 & 1.567\tablenotemark{a}\\
H$_2$ 0--0 S(3)   &\phn9.6649 & 1.055 $\pm$ 0.288\\
$[$\ion{Ne}{ii}]  &12.8136 & 6.972 $\pm$ 0.252\\
$[$\ion{Ne}{iii}] &15.5551 & 2.666 $\pm$ 0.168\\
H$_2$ 0--0 S(1)   &17.0348 & 1.808 $\pm$ 0.163\\
$[$\ion{S}{iii}]  &18.7130 & 2.248 $\pm$ 0.119\\
\hline
$[$\ion{Fe}{ii}]  &\phn5.3402 & $<$ 1.92\\
H$_2$ 0--0 S(7)   &\phn5.5112 & $<$ 1.76\\
H$_2$ 0--0 S(4)   &\phn8.0250 & $<$ 1.53\\
$[$\ion{Ar}{iii}] &\phn8.9914 & $<$ 1.03\\
$[$\ion{S}{iv}]   &10.5105 & $<$ 0.60\\
H$_2$ 0--0 S(2)   &12.2786 & $<$ 0.96\\
$[$\ion{Ne}{v}]   &14.3217 & $<$ 0.41\\
$[$\ion{Fe}{ii}]  &17.9360 & $<$ 0.34\enddata
\tablenotetext{a}{We quote here the measurement of Pf$\alpha$ 
intensity provided by the fitting procedure,
however its value is unreliable because it falls in the 
wing of the broad PAH 7.7 $\mu$m emission feature.}
\end{deluxetable}


\subsubsection{Constraints on AGN excitation}
\label{sec:sf_agn}

Nebular emission lines provide a powerful diagnostic tool to
determine the nature of the source of energy powering the
emission. Classically, the determination is done
using optical emission lines. Optical spectra of the nuclear
region of NGC~4435 were observed by \citet{hofi95}. Using the
diagnostic diagrams of \citet{veil87}, \citet{hofi97} 
classified NGC~4435 as a transition object between LINER
and \ion{H}{ii}, via the value of [\ion{O}{i}]6300/H$\alpha$
(=0.13) which, however, has an uncertainty of 30-50\%. Revisiting
the classification using the AGN/\ion{H}{ii} separation proposed by \citet{kewl01}, 
the [\ion{O}{i}]6300/H$\alpha$ ratio is only marginally larger
than that expected for \ion{H}{ii} regions.

The IRS spectrum allows us to make use of the ratio
[\ion{Ne}{iii}]15.5/[\ion{Ne}{ii}]12.8 which is a strong infrared diagnostic for
the hardness of the ionizing source. We compared the observed
ratio of [\ion{Ne}{iii}]15.5/[\ion{Ne}{ii}]12.8=0.38 with CLOUDY models
\citep{ferl98}. We computed several photoionization models using as
ionizing source the spectrum of a 35000 K star (with Kurucz
atmosphere model) plus a $10^5$ K blackbody, representing the AGN
contribution \citep[a similar test was done by][]{lutz98}.
Assuming an ionization parameter $\log(U)=-2$, and n$_{\rm{H}}=70$
(derived from \ion{S}{ii} optical lines), the observed ratio
[\ion{Ne}{iii}]15.5/[\ion{Ne}{ii}]12.8 limits the AGN contribution to the ionizing
flux to be less than 2\%.


\subsubsection{H$_2$ rotational emission}
In Table \ref{tab:emlines}  we provide
the  intensity of detected H$_2$ rotational emission lines.

\begin{figure}
\plotone{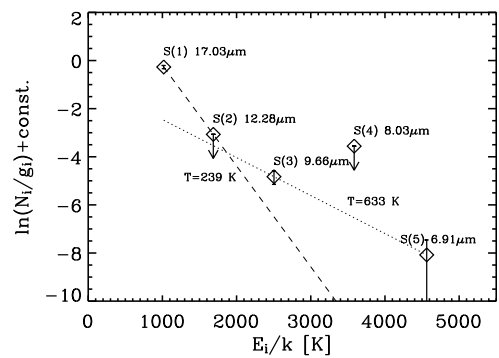}
\figcaption{Excitation diagram for H$_2$ rotational lines observed in the NGC~4435 IRS spectrum.
Diamonds show the natural logarithm of the number of molecules in the state $i$ computed with
eq. \ref{eq:nilum} divided by the statistical weight
plotted versus the energy of the excited state with respect to
the ground state. Error bars represent 1$\sigma$ errors on line intensity, while arrows show
upper limits on H$_2$ S(4) and H$_2$ S(2) lines. The dashed and dotted lines show the fits to
the warm and hot components, respectively.}
\label{fig:h2diag}
\end{figure}

The first excited rotational state of H$_2$ that can provide a transition toward the
fundamental level has an angular moment $J=2$ and an energy of 510 K, thus the
H$_2$ does not radiate for temperatures typical of the cold molecular medium
(some 10 K) but only for temperatures above 100 K.
The investigation of the state of warm molecular gas can be done using the
excitation diagram, shown in Figure \ref{fig:h2diag}, where the
natural logarithm of the number of molecules in the state $i$
($N_i$) divided by the statistical weight of the state ($g_i$) is
plotted against the energy ($E_i$) of the excited state with respect to
the ground state.
The number of molecules in state $i$ is given by (assuming
a density smaller than the critical density):

\begin{equation}
\label{eq:nilum}
N_i=\frac{L_i}{A_i h \nu_i}
\end{equation}
where $L_i$ is the luminosity of the line emitted by the transition
starting from the state $i$, $A_i$ is the transition probability, and
$\nu_i$ the frequency of the transition.

Assuming local thermal equilibrium, the ratio between
number of molecules in state $i$ and the total number of H$_2$
molecules per unit of volume is given by the Boltzmann equation
\citep{parm91}:

\begin{equation}
\frac{n_i}{n_{\rm H_2}}=g_i\frac{hcB}{2kT}\exp \left( -\frac{E_i}{kT}\right)
\label{eq:boltz}
\end{equation}
where $B$ is the H$_2$ rotational constant.

For an isothermal gas all the data in the excitation diagram of Figure
\ref{fig:h2diag} should lay on a linear relation $y=a+b x$ with
$b=-1/T$. However, the observed values do not lay on the same line,
suggesting the presence of two (or more) components at different
temperatures, a ``warm'' component at $T \le 240$ K traced by the
S(1) line (and the S(2) upper limit) and a ``hot'' component at $T
\simeq 630$ K, traced by lines at shorter wavelengths. With equations
\ref{eq:nilum} and \ref{eq:boltz} we can estimate the mass of
the ``warm'' and ``hot'' components.
Assuming a temperature of 200 K for the warm component, we find a
mass of $6.3\times 10^4 M_\odot$, while for the ``hot''
component we find a mass of $7.1\times 10^3 M_\odot$.

The excitation mechanism for hot and warm molecular gas in galaxies is commonly attributed to
UV flux from young stars in photodissociation regions (PDR),  shocks or
X-ray flux from an active nucleus \citep{burt92,rigo02}.

The upper limit on the temperature derived from S(1) and S(2)
lines excludes AGN excitation. In fact \citet{rigo02}
find that in Seyfert galaxies the temperature from these lines is
typically around 350 K. Indeed, the temperature we find is
compatible with PDR models by \citet{burt92}. Finally, it is
worth noting that the mass of the warm gas is only a tiny fraction
($\sim 0.06\%$) of the total molecular gas observed in NGC~4435
from CO lines \citep[$\sim 10^8 M_\odot$,][]{voll05}. This
fraction is also significantly smaller than that found in typical
powerful starbursts (1 -- 10\%) or in Seyfert galaxies \citep[up to 30\%,][]{rigo02}. 
This is consistent with our finding
that the star formation in NGC~4435 has already declined and is
more characteristic of a post-starburst phase, as we discuss in
the following sections.
 
Note that the above values are not corrected for dust attenuation. However,
the attenuation we derived from the model fit to the global SED in the following
sections is quite low ($A_\lambda \sim 0.1 - 0.2$) at these wavelengths. 


\subsubsection{Abundance determinations}
\label{sec:abbdet}
From the intensity of the Pf$\alpha$, neon and argon
lines we could obtain an estimate of the metallicity of the ionized gas,
following the procedure outlined in \citet{verm03}.
In fact, for a generic element X and as long as the current
density is lower than the critical density for collisional
de-excitation of the involved transitions, the abundance relative to
hydrogen is given by:
\begin{equation}
\frac{\rm X}{\rm H} = \frac{j_{\rm Pf\alpha}}{
F_{\rm Pf\alpha}}\left(
\frac{F_{\rm [XII]}}{q_{\rm [XII]}h\nu_{\rm [XII]}}+
\frac{F_{\rm [XIII]}}{q_{\rm [XIII]}h\nu_{\rm [XIII]}}+....\right)
\label{eq:abund}
\end{equation}
where $j_{\rm Pf\alpha}$ is the emissivity for the Pf$\alpha$, and
$q_{\rm [XII]}$ is the collisional excitation rate of [\ion{X}{ii}]
transition (for electronic density near to zero). 
The above abundances can be translated into metallicity estimations
by dividing them by standard ISM abundances. 

We notice however that, while we are confident on the determination of metal
line intensities, the Pf$\alpha$ falls in the wing of a broad PAH emission feature.
The measurement of Pf$\alpha$ intensity is thus unreliable, even if the fitting procedure
gives a small uncertainty (reported in table \ref{tab:emlines}). 
For this reason we do not proceed with a direct determination of the abundances in
this way. However an alternative way to exploit the good
measurements of metallic lines in order to derive the metallicity is to use the 
hydrogen line intensity as predicted from the  fit to the global SED of the galaxy. This will be
discussed in a following section.


\section{Analysis of the panchromatic SED}
\label{sec:sedglobal}

The analysis of the MIR spectrum  made in
the previous section indicates that a star formation activity
is present in the circumnuclear disk of NGC~4435. 
In order to better characterize the emission
mechanisms and the stellar population content of this galaxy we will
model the panchromatic SED of NGC~4435 by means of our
spectro-photometric code GRASIL\footnote{Available at
http://adlibitum.oats.inaf.it/silva/default.html or
http://web.oapd.inaf.it/granato/grasil.html} 
with the recently updated treatment of PAH emission based on ISO data
\citep[for details on GRASIL see][]{silv98,silv99,gran00,bres02,panu03,panu05,vega05}.

The panchromatic SED was constructed by combining the IRS spectra
(including the main nebular lines) with archival data
matching as far as possible the IRS apertures.
J, H and Ks fluxes within an aperture of 5\arcsec\ radius were
taken from the 2MASS database at IRSA\footnote{The NASA/IPAC Infrared 
Science Archive is operated by
the Jet Propulsion Laboratory, California Institute of Technology,
under contract with the National Aeronautics and Space
Administration.}. 
IRAC fluxes in the 3.6, 4.5, 5.8, and 8.0~$\mu$m bands were 
extracted in the same aperture from archival images as 
explained in section \ref{irac}. IRAS 60, and 100~$\mu$m fluxes (obtained
from NED) were used to define the FIR SED together with MIPS total fluxes. 
Although the apertures of MIPS and IRAS are larger than
the IRS aperture, the above fluxes are dominated by emission from
the dusty circumnuclear disk as clearly shown by the 24~$\mu$m MIPS image. 
This is not true, however, at 12 $\mu$m, where we expect a significant contribution
from the old stellar population; thus the 12 $\mu$m IRAS flux has not been included in the
SED. We also excluded the IRAS 25~$\mu$m flux (0.22 Jy, from NED) because 
it is not compatible with the 24~$\mu$m MIPS flux; this is probably due to an
incorrect background subtraction due to the poor resolution of IRAS.
Radio data were taken from \citet{naga00}, \citet{wrob91}, \citet{beck91}, \citet{dres78}, 
and the FIRST-VLA survey. 
The SED is completed with GALEX UV fluxes
extracted within a 5\arcsec\ aperture (Boselli \& Cortese, private communication).
The broad band fluxes are given in Table \ref{tab:data}.

\begin{deluxetable}{rcccl}
\tablewidth{0pt}
\tablecaption{Broad band fluxes of NGC~4435  applicable
for 5\arcsec\ central region\label{tab:data}}
\tablehead{
\colhead{ $\lambda$ }& 
\colhead{Band /}& 
\colhead{Flux}&
\colhead{Uncert.}&
\colhead{Ref.}\\
\colhead{($\mu$m)}& 
\colhead{Instrument}&
\colhead{(Jy)}&
\colhead{(mJy)}&
\colhead{}
}
\startdata
1.25  &J      &0.11     &12.1   &1\\
1.65  &H      &0.14     &15.3   &1\\
2.17  &Ks     &0.12     &13.2   &1\\
3.6   &IRAC   &0.056    & 5.6   &2\\
4.5   &IRAC   &0.033    & 3.3   &2\\
5.8   &IRAC   &0.039    & 3.9   &2\\
8.0   &IRAC   &0.062    & 6.2   &2\\
23.7  &MIPS   &0.11     & 22.0  &2\\
60    &IRAS   &1.99     &597.0  &3\\
71.0  &MIPS   &2.27     &454.0  &2\\
100   &IRAS   &4.68     &1404.0 &3\\
156   &MIPS   &3.7      &740.0  &2\\
20000 &VLA    &$<$0.0011& -     &4\\
60000 &VLA    &$>$0.0012& -     &5\\
61856 &       &$<$0.007 & -     &6\\
126050&       &$<$0.004 &-      &7\\
200000&VLA    &0.0022   &0.216  &8\enddata
\tablerefs{(1) IRSA; (2) this work;
(3) NED; (4) Nagar et al. 2000; 
(5) Wrobel \& Heeschen 1991; (6) Becker et al. 1991; (7) Dressel\& Condon 1978; (8) FIRST VLA survey.}
\end{deluxetable}


\subsection{The models}
\label{sec:grasil}
Since the broad band SED obtained in the previous section 
is produced by the star-forming circumnuclear disk and the old
population, including also the stars seen in projection,
we adopted a simple composite model for the fit:
a starburst plus an old stellar population.
The old stars are modelled as an unattenuated simple stellar population
(SSP) with given age and metallicity. 
The SSPs are based on Padova stellar models and
include the contribution from
dusty circumstellar envelopes around asymptotic giant branch stars.
This fact may be relevant because these old stars 
leave an excess around 10 $\mu$m \citep{bres98,bres01,bres06}. 

The starburst contribution was modelled with GRASIL assuming
different values for the age ($t_{\rm{b}}$), metallicity and star formation history.
The latter was assumed as an exponentially decreasing SFR parameterized by 
the e-folding time, $\tau_{\rm{b}}$.
We consider two phases for the ISM: the molecular clouds (MCs) which surround
young star clusters and the diffuse component. 
For the distribution of both the diffuse ISM and stars we adopted a disk geometry,
with an inclination of $45^\circ$ with respect to the line-of-sight \citep[as suggested by
observations, e.g.][]{cocc06}. The density of dust and stars is
assumed to be exponential in both radial and vertical
direction, with different scales for dust and stars
\citep[see][for more details]{silv98}.

Among the GRASIL physical quantities that mostly affect the shape of the SED
we recall 
{\em (i)} the total gas mass $M_{\rm gas}$,
{\em (ii)} the fraction of molecular mass\footnote{In the GRASIL model we identify
the molecular gas with the clouds surrounding young star
clusters. Molecular gas could also be present in the diffuse component but
the code doesn't differentiate between atomic and molecular gas in this phase.}
to total gas mass ($M_{\rm{mol}}/M_{\rm{gas}}$),
{\em (iii)} the optical depth of molecular clouds at $1 \mu$m ($\tau_{\rm{1}}$), and
{\em (iv)} the escape time-scale of newly born stars from their parent MCs
($t_{\rm{esc}}$).
The dust composition and optical properties are set to
reproduce the extinction and
emission properties observed in the diffuse ISM of our Galaxy.
The gas-to-dust mass ratio ($G/D$) is set proportional to 
the metallicity ($Z$), with $G/D=110$ for solar metallicity.

For the nebular lines we adopted the large \ion{H}{ii} region
library computed by \citet{panu03}.
Finally, the adopted IMF is a Salpeter IMF \citep{salp55}
in the mass range from 0.1 to 100 $M_\odot$.

\begin{deluxetable}{ccccc}
\tablewidth{0pt}
\tablecaption{Parameters for the burst model library.\label{tab:granew}}
\tablehead{
\colhead{$\tau_{\rm{b}}$}& 
\colhead{$\log(t_{\rm{b}}$)}& 
\colhead{$t_{\rm{esc}}$}&
\colhead{$M_{\rm{mol}}/M_{\rm{gas}}$}&
\colhead{$\tau_{1}$}\\
\colhead{(Myr)}& 
\colhead{(yr)}&
\colhead{(Myr)}&
\colhead{}&
\colhead{}
}
\startdata
 10 - 100 &  6.0 - 8.8 &   1 - 99   & 0.10 - 0.99&0.01 - 180
\enddata
\end{deluxetable}


\subsection{Results of the SED fit}
\label{sec:resultsed}

\begin{deluxetable}{cccccc}
\tablewidth{0pt}
\tablecaption{Best fit parameters for the burst model.\label{tab:sedparam}}
\tablehead{
\colhead{$\tau_{\rm{b}}$}& 
\colhead{$t_{\rm{b}}$}& 
\colhead{$Z$}&
\colhead{$t_{\rm{esc}} $}&
\colhead{$M_{\rm{mol}}/M_{\rm{gas}}$}&
\colhead{$\tau_{\rm{1}}$}\\
\colhead{(Myr)}&
\colhead{(Myr)}&
\colhead{(Myr)}&
\colhead{}&
\colhead{}&
\colhead{}
}
\startdata
55&186&0.02&3.0 &0.20 &0.25\enddata
\end{deluxetable}

\begin{deluxetable*}{cccccccc}
\tablewidth{0pt}
\tablecaption{Derived quantities of the SED models.\label{tab:sedderiv}}
\tablehead{
\colhead{$M_{\star \rm F}^{\rm{SB}}$}& 
\colhead{$M_{\rm gas}$}& 
\colhead{$M_{\rm old}^{5\arcsec}$}&
\colhead{$L_{\rm SB}^{5\arcsec}$}&
\colhead{$L_{\rm old}^{5\arcsec}$}&
\colhead{$\langle \rm SFR\rangle$}&
\colhead{SFR$_{\rm{c}}$}&
\colhead{$A_{\rm{V}}^{\rm{mod}}$}\\
\colhead{($M_\odot$)}&
\colhead{($M_\odot$)}&
\colhead{($M_\odot$)}&
\colhead{(\%)}&
\colhead{(\%)}&
\colhead{($M_\odot$ yr$^{-1}$)}&
\colhead{($M_\odot$ yr$^{-1}$)}&
\colhead{(mag)}
}
\startdata
$1.40\times10^8$&$1.36\times 10^{8}$&$8.2\times 10^{9}$&32.0&68.0&0.75&0.089&2.69\enddata
\end{deluxetable*}

\begin{figure*}
\plottwo{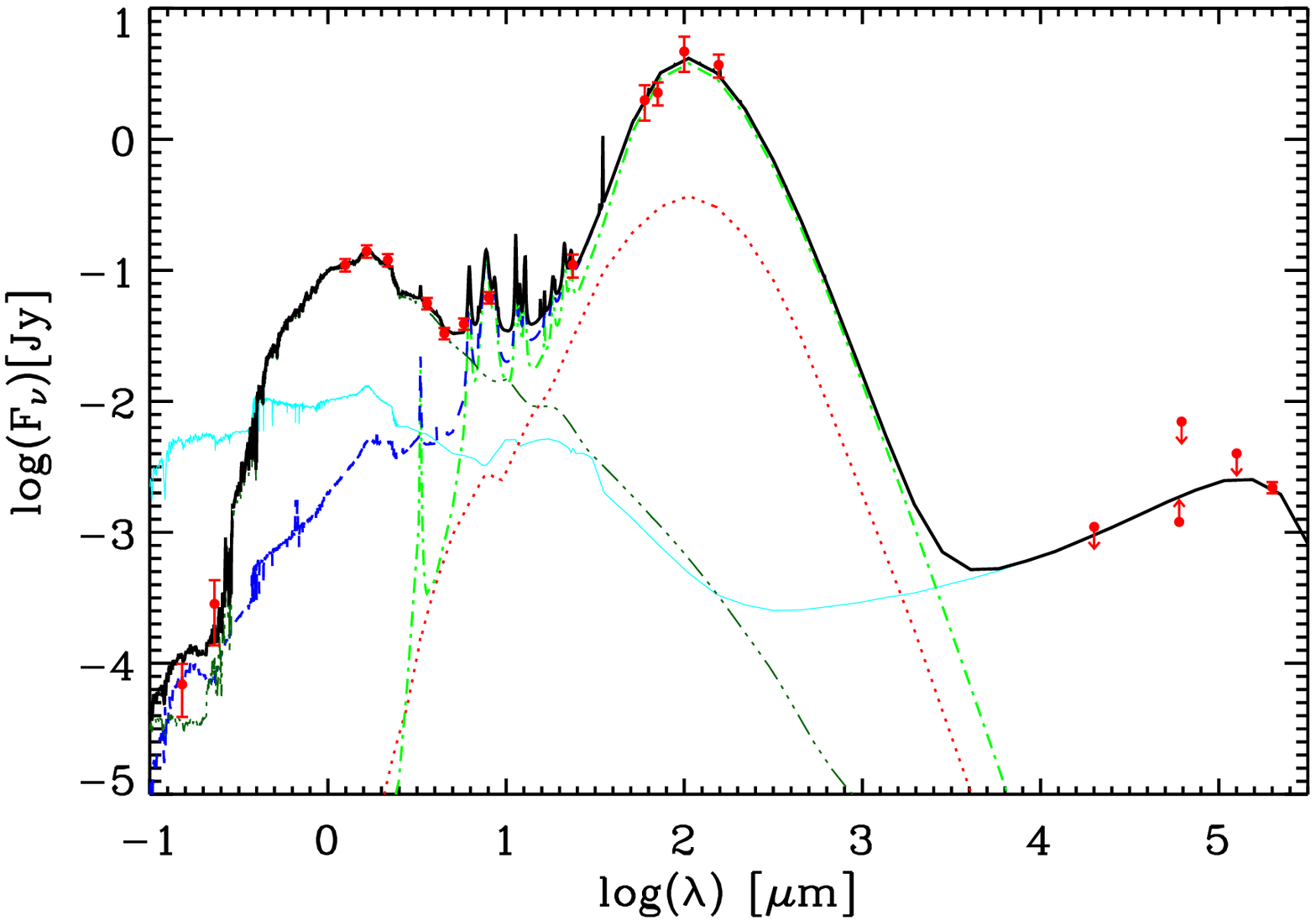}{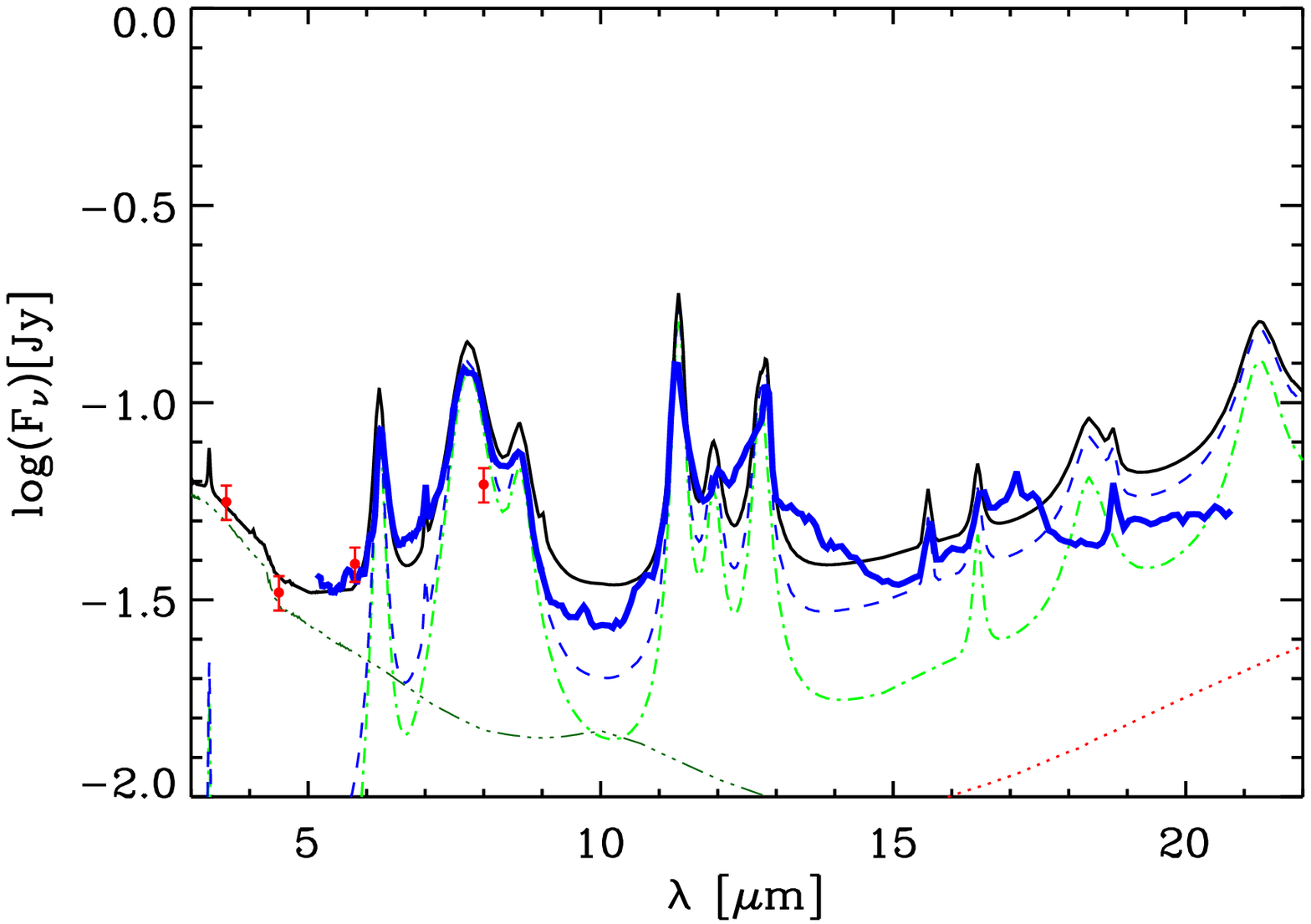}
\figcaption{Comparison between the observed SED of 
the central region of NGC~4435 and our model. The thick solid black line
represents the model for the total SED, i.e. the starburst
component plus the old stellar component; the three dots-dashed darkgreen line
represents the contribution from the old stellar population, and
the dashed blue line represents the total contribution from the burst
of star formation, the dotted red line represents the emission from
molecular clouds. The dot-dashed green line represents the diffuse medium
emission and the thin cyan solid line  denotes the emission from
stars of the starburst component without applying the extinction from dust. The
red-filled circles are the broad band data. 
\textit{Left}: Comparison from 0.1$\mu$m to 100 MHz.
\textit{Right}: 
Comparison for the MIR wavelengths.
The thickest blue solid line represents the IRS \emph{Spitzer} spectrum.
} \label{fig:central_sed}
\end{figure*}

We computed a large grid of simulated SED with GRASIL. In Table \ref{tab:granew}
we list the ranges of values used for the library. Among these models, we identified
the best fitting model as the one that maximize a likelihood function.
The likelihood function was built using UV, MIPS, IRAS and radio  data and their uncertainties quoted
in Table 2 and the intesities of observed line H$\alpha$, H$\beta$ \citep[from][]{hofi97}
and [\ion{Ne}{ii}]12.8$\mu$m. The best fit model is shown in Figure \ref{fig:central_sed}, where
the left panel shows the panchromatic fit, from the UV to radio wavelengths,
while the right panel is a zoom into the MIR spectral region.
The corresponding parameters and some important derived quantities are displayed
in Tables \ref{tab:sedparam} and \ref{tab:sedderiv} respectively.

The wide wavelength coverage of the SED imposes strong constraints
to the model, because different spectral regions
probe the age sequence of the stellar populations.
Broadly speaking, nebular emission lines are dominated by massive stars,
radio emission is contributed by stars younger than about 50 Myr while
the FIR SED includes also a significant contribution from intermediate mass stars.
The UV/FIR ratio implies that the burst luminosity is almost completely
reprocessed by dust and emitted in the FIR, thus the FIR luminosity fixes the
bolometric luminosity of the burst. The emission line luminosities and radio fluxes 
constrain the star formation rate in the last 50 Myr determining the timescale 
$\tau_{\rm b}$. Once the current SFR and $\tau_{\rm b}$ are determined, the 
age of the burst is fixed by its luminosity.
Finally, the H$\alpha$ and H$\beta$ constrain 
the attenuation at optical wavelenght (see section \ref{sec:attlaw} for more details).

Our assumption of an exponentially declining SFR is 
the simplest model consistent with a Schmidt-type law 
that reproduces the multiwavelength SED.
Other more complex scenarios could probably be able to fit the data 
introducing, however, further free parameters that are not
justified.
Indeed, stars in the obscured burst older than $\sim$ 50 Myr do not 
provide observable ``features'' to disentangle
the detailed shape of the star formation history. 

It is also interesting to consider the effects
of a underlying star formation activity already present in
NGC~4435 before the interaction (i.e. older than $\sim$ 200 Myr).
Stars originated from this activity would unlikely remain
confined to the dusty disk. For example in our galaxy,
stars older than 150 Myr show larger spatial distribution
than younger stars \citep{robi03,panu06}.
Thus the intermediate age stars would now be mostly outside
the dusty disk and, as a consequence, they would suffer from
only a small or negligiable attenuation.
Since stars 200--500 Myr old are still important emitters at
UV wavelengths, they would contribute significantly to the
GALEX fluxes. However, the measured UV GALEX and
NIR fluxes within the 5\arcsec\ aperture (Fig. \ref{fig:central_sed} left panel), are
already entirely accounted for by the old population and the
obscured starburst.
Actually, the observed FUV flux within the 5\arcsec\ aperture
corresponds to an activity not larger than SFR $\sim 3\times 10^{-4} M_\odot$  yr$^{-1}$
between 200 and 500 Myr ago.
This suggests that no significant underlying intrinsic star formation 
activity was present and,
consequently, that the observed burst was
triggered by the interaction.

The  old stellar population dominates the emission in the optical and NIR range
but is significant also in the MIR. 
In fact we see from Figure \ref{fig:central_sed} that the fractional contribution of
the starburst to the IRS spectrum increases with wavelength. 
This is consistent with the results of the spatial analysis along the slit
presented in section \ref{sec:irs_redu}. 
We find that the continuum below $\sim 6$
$\mu$m looks extended, while the emission at longer wavelengths (in particular the PAH
features) looks unresolved, in agreement with the different concentration observed in 
ISO LW2 and LW3 images \citep{bose03}, IRAC and MIPS images, as detailed below.

Finally we notice that the fit of the SED does
not require an AGN component, which adds further support for the lack
of AGN activity in NGC~4435.

In the following we discuss in more detail the results of the fit.


\section{Discussion}
\label{sec:discussion}


\subsection{The star formation history and starburst luminosity}

The old stellar component is well fitted by 
a solar metallicity SSP of 
8 Gyr and a mass of $8.2\times10^9~M_\odot$.  
Note that the value here derived for the stellar mass depends on the adopted IMF. 
\citet{cocc06} give an estimate of the mass to light ratio ($M/L$) of the population
that can be directly compared with our old stellar population. Using 
the $I$ band they measured $M/L(I)=2.2\cdot M_\odot/L_\odot(I)$, while for our old 
stellar population it is $M/L(I)=2.98\cdot M_\odot/L_\odot(I)$. The discrepancy
in the $M/L$ can be solved by changing the slope of the 
IMF at lower masses \citep[as in the Kroupa IMF,][]{krou98} or adopting a higher lower mass limit
(e.g. 0.2 $M_\odot$) without changing the predicted SEDs.

The nuclear disk is in a post-starburst phase
($\tau_{\rm{b}}= 55$ Myr and $t_{\rm{b}}$ = 186 Myr) 
with a current star formation rate
(SFR$_{\rm{c}}=0.089~M_\odot$~yr$^{-1}$) about one order of magnitude
smaller than the average star formation rate 
($\langle {\rm SFR}\rangle= 0.75~M_\odot$~yr$^{-1}$).
The quality of radio data, however, prevents a determination of 
$\tau_{\rm b}$ better than $\pm 20$\%; this results in an uncertainty on 
the age of the sturburst $t_{\rm{b}}$ of the same order.
The mass of gas converted into stars during the burst is 
$M_{\rm{\star F}}^{\rm{SB}}=1.40\times10^8~M_\odot$; however with the adopted IMF, only $\sim$
80\% still survive. The star forming component emits 32\%
of the bolometric luminosity in the central 5\arcsec\ of the galaxy, 
while accounting for 1.4\% of the stellar mass.
Most of the starburst luminosity  is provided by A stars, reprocessed
by the diffuse dust and re-emitted in the 8-1000~$\mu$m spectral region
($L_{\rm IR}\simeq 5.5\times 10^{35}$~W).

A population of A stars in the central  $\sim 550$ pc has already been detected
by optical observations \citep{kenn95}.
More recently, \citet{sarz05} estimated 
that the contribution of the young burst from 3000~\AA \ to 5700~\AA\ in the nuclear region
(0\farcs13 radius) is around 12\%.
In the same spectral region we estimate that the 
fractional contribution of the starburst is only 6\%.
It is however hard to compare the two estimates because
our value refers to a much larger region (5\arcsec).
Indeed it is sufficient that in the very central region
the attenuation be reduced \citep[as estimated by][]{sarz05} to
eliminate this discrepancy without affecting significantly the
constraints from the other spectral bands.

The strongly declining SFR makes this galaxy 
a typical cases where different SFR indicators, probing different 
timescales, provide estimates that are apparently discrepant.
In fact with the classical SFR estimation based on 
the IR luminosity  provided by \citet{kenn98}
we get SFR$_{\rm{IR}}= 0.25~M_\odot$~yr$^{-1}$, in between the
average SFR and the current one.
Using the 1.4 GHz flux with the \citet{panu03} calibration we get
SFR$_{\rm{1.4GHz}}\simeq 0.117~M_\odot$~yr$^{-1}$. This value 
is more similar to the current SFR value than the IR based estimate
because the radio emission is due to free-free emission from HII
regions and synchrotron emission from core collapse supernovae.


\subsection{The geometry and the attenuation law}
\label{sec:attlaw}

Though the star formation history is well constrained by the fit of the total SED,
a further refinement to the relative spatial distribution of stars and dust
can be obtained by  the comparison of the predicted and observed
H$\alpha$ and H$\beta$ line intensities.
It was shown in previous works that a system with
an equal spatial distribution of stars and dust produces a much smaller reddening
than a ``screen'' geometry \citep[e.g.][]{witt00}. 
With the assumption of equal scales for stars and dust
and given the strong attenuation implied by the observed UV/FIR ratio,
the H$\alpha$ intensity predicted  by our best fit model would be much smaller than
the observed value \citep[$I$(H$\alpha$)= 2.82$\times 10^{-17}$~W m$^{-2}$,][]{hofi97}
and the predicted gas mass
about twice the quoted value $\sim 10^8M_\odot$ \citep{voll05}.
To solve this discrepancy we need to decrease the attenuation in the optical band
leaving unchanged that in the UV.
Without modifying the optical properties of dust grains, this can be achieved
by assuming a geometry where the diffuse dust is more extended than
the stellar distribution. This has the effect of
increasing the reddening, because
it reddens the attenuation law making it more similar to the extinction of
a uniform slab. Adopting a radial (vertical) scale length 
of 160 (22) pc for the diffuse dust and of
70 (14) pc for the stellar component
the predicted intensities for H$\alpha$ and H$\beta$ are $I$(H$\alpha$)= 2.70$\times 10^{-17}$
W m$^{-2}$, and $I$(H$\beta$)= 4.99$\times 10^{-18}$
W m$^{-2}$ well in agreement with the observed value \citep[4.99$\times 10^{-18}$
W m$^{-2}$ for H$\beta$ from][]{hofi97}.
The required dust mass is 
$1.23 \times 10^{6}~M_\odot$. For a solar metallicity,
as predicted by other observables discussed below,
this implies a gas mass  $M_{\rm{G}}\sim 1.36\times 10^8~M_\odot$, 
fairly near to the observational value.

\begin{figure}
\plotone{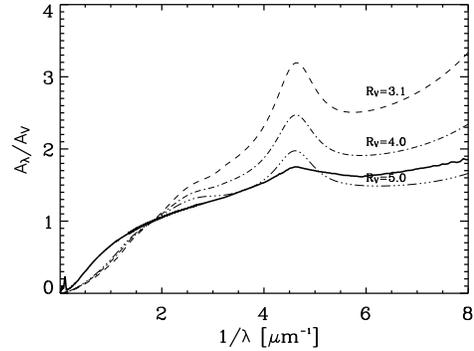}
\figcaption{Comparison of Cardelli extinction curves with different
R$_{\rm{V}}$ with the attenuation curve derived from the best fit
model (solid line). \label{fig:attenuation}}
\end{figure}

The estimated mean attenuation of the system in the V band is 
$A_{\rm{V}}^{\rm{mod}}=2.69$ mags. 
The resulting attenuation law is shown in Figure \ref{fig:attenuation}. 
In the same figure we plot, for comparison,
three  different  extinction curves from \citet{card89}. 
We stress that the dust properties adopted in the model are such that
they reproduce the Galactic extinction law, corresponding to
$R_{\rm{V}}$=3.1. Thus the geometry of the model 
(including the effects of age dependent
attenuation) produces an effective attenuation law which
is very different from the adopted intrinsic extinction properties of the
dust \citep{gran00,panu03,panu06}.


\subsection{The gas and star metallicity}

The best fit to the NGC~4435 SED is obtained by adopting solar metallicity for
both stars and gas in the starburst. 

Following the detailed study of nebular emission models of \citet{panu03}
we can also compute the predicted intensities of selected
emission lines. This is particularly useful for abundance determinations
with infrared lines in the absence of good measurements of hydrogen recombination
lines \citep[see][]{panu03}, as in the present case. In fact from the SFR obtained
from the fit of the SED, GRASIL also computes the ionizing flux and thus,
the observed intensity of the main emission lines. This can be considered as
an alternative method to obtain the ionizing flux from a direct measurement
of the hydrogen recombination lines. Adopting solar metallicity and standard ISM abundances
we obtain the following predicted intensity: $I($Pf$\alpha)=3.46\times 10^{-18}$ W m$^{-2}$,
$I($[\ion{Ne}{ii}]$)=7.01\times 10^{-17}$ W m$^{-2}$,
$I($[\ion{Ne}{iii}]$)=2.60\times 10^{-17}$ W m$^{-2}$, 
$I($[\ion{Ar}{ii}]$)=3.17\times 10^{-17}$ W m$^{-2}$ and
$I($[\ion{S}{ii}]$)=1.71\times 10^{-17}$ W m$^{-2}$. 
For the metallic lines these figures are in good
agreement with the observed values reported in Table \ref{tab:emlines}, thus
providing an independent confirmation that the gas metallicity is solar.
The value found for Pf$\alpha$ is much lower than that derived from the fit of
the line, suggesting that the blend with PAH prevents an accurate
measurement (see section \ref{sec:abbdet}).


\subsection{PAHs and IRS continuum}

The left panel of Figure \ref{fig:central_sed} shows a detailed
comparison MIR \emph{Spitzer} spectrum and the model. The continuum
and PAH features up to 13~$\mu$m are well fitted by the starburst
model. At wavelengths longward of 13~$\mu$m, while the predicted
continuum matches fairly well that observed, the predicted PAH
features are in disagreement with those observed, both in
central wavelength and strength. Only the 16.4~$\mu$m
feature is reproduced.

We recall that our model for PAH emission is based on the \citet{lidr01}
model, which predicts PAH features at 3.3, 6.2, 7.7, 8.6, 11.3,
11.9, 12.7, 16.4, 18.3, 21.2, and 23.1 $\mu$m. The model of PAH
features  at $\lambda < 13~\mu$m was based on ISO data for the
diffuse galactic medium. While the FWHM and the cross sections for
the features at $\lambda > 13$ $\mu$m
were taken from the theoretical study of 40
families of PAHs driven by \citet{mout96}, because of the lack of
corresponding observational data.

More recently, ISO extended the observed wavelength range, and
confirmed the feature at 16.4 $\mu$m, and also found new features
at 14.2 and 17.4 $\mu$m, as well as a weak
plateau between 15 and 20 $\mu$m \citep{mout00,stur00,vank00}. 
This plateau is attributed to a
blend of many emission features provided by the interstellar PAH
families present in the sources, and its variation can be
accommodated by variations in the PAH populations \citep{peet04a}. 
Lately, the higher sensibility of
\emph{Spitzer} confirmed the existence of PAH features at 14.2,
16.4 and 14.7 $\mu$m, and a new and prominent feature at 17.1
$\mu$m, among other weaker ones \citep[e.g.][]{bran04,smit04,wern04}.
However, the lack of features at 18.3 $\mu$m, 21.2 $\mu$m, and
23.1 $\mu$m has  been confirmed.

The disagreement between model and data was already noticed by
\citet{peet04b}, \citet{smit04} and \citet{dale05}
who, on the other hand,  found very good agreement between the Li
\& Draine model and the data for $\lambda < 13$ $\mu$m. They
attributed the disagreement at wavelengths longward of 13 $\mu$m
to the large PAH sensitivity to their global molecular structure,
since they involve the motion of the molecules as a whole, and
therefore, they are very dependent on the exact PAH species in the
emitting source.

Work is in progress to improve our PAHs emission model
based on new \emph{Spitzer} observations.


\begin{figure*}
\plottwo{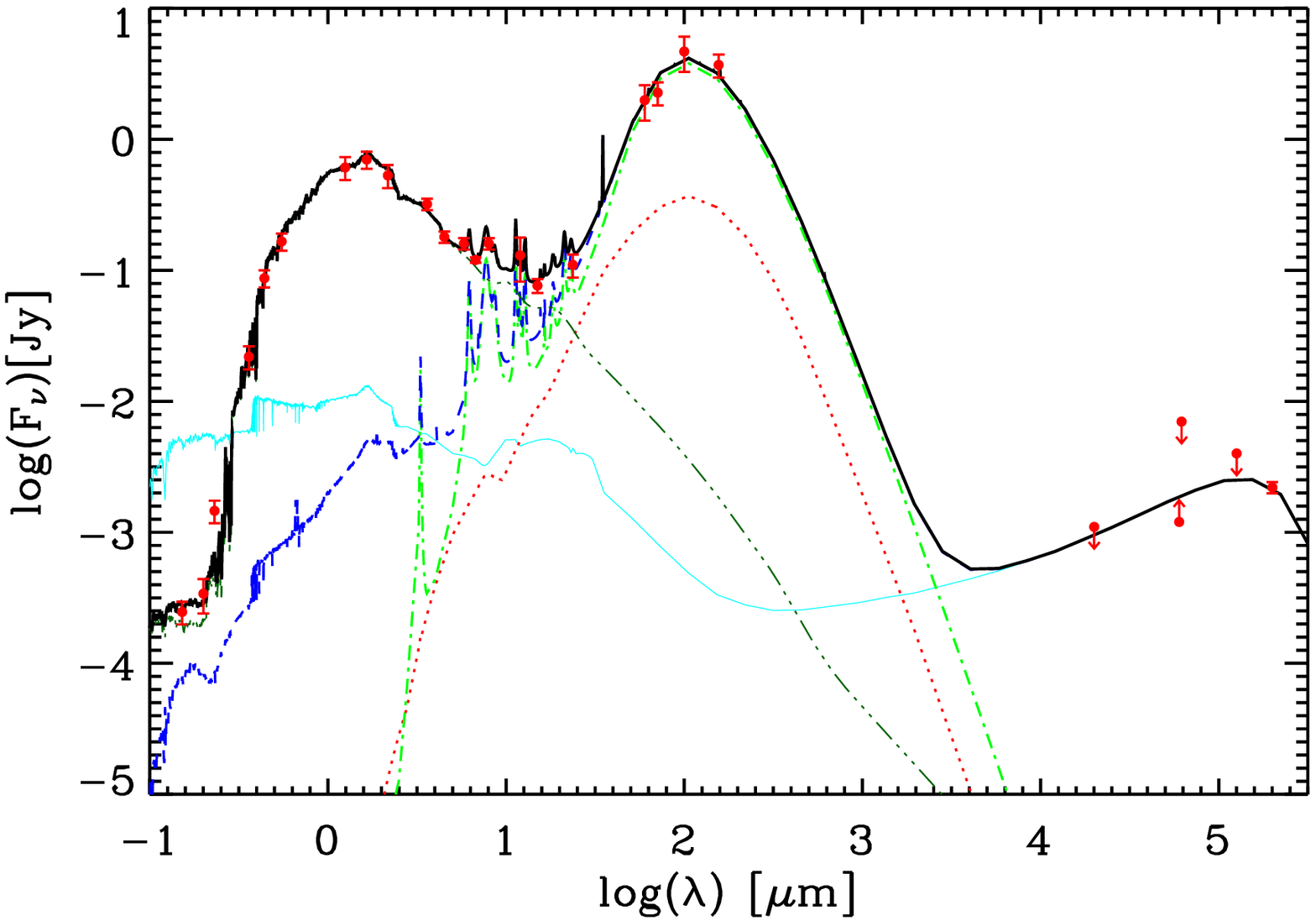}{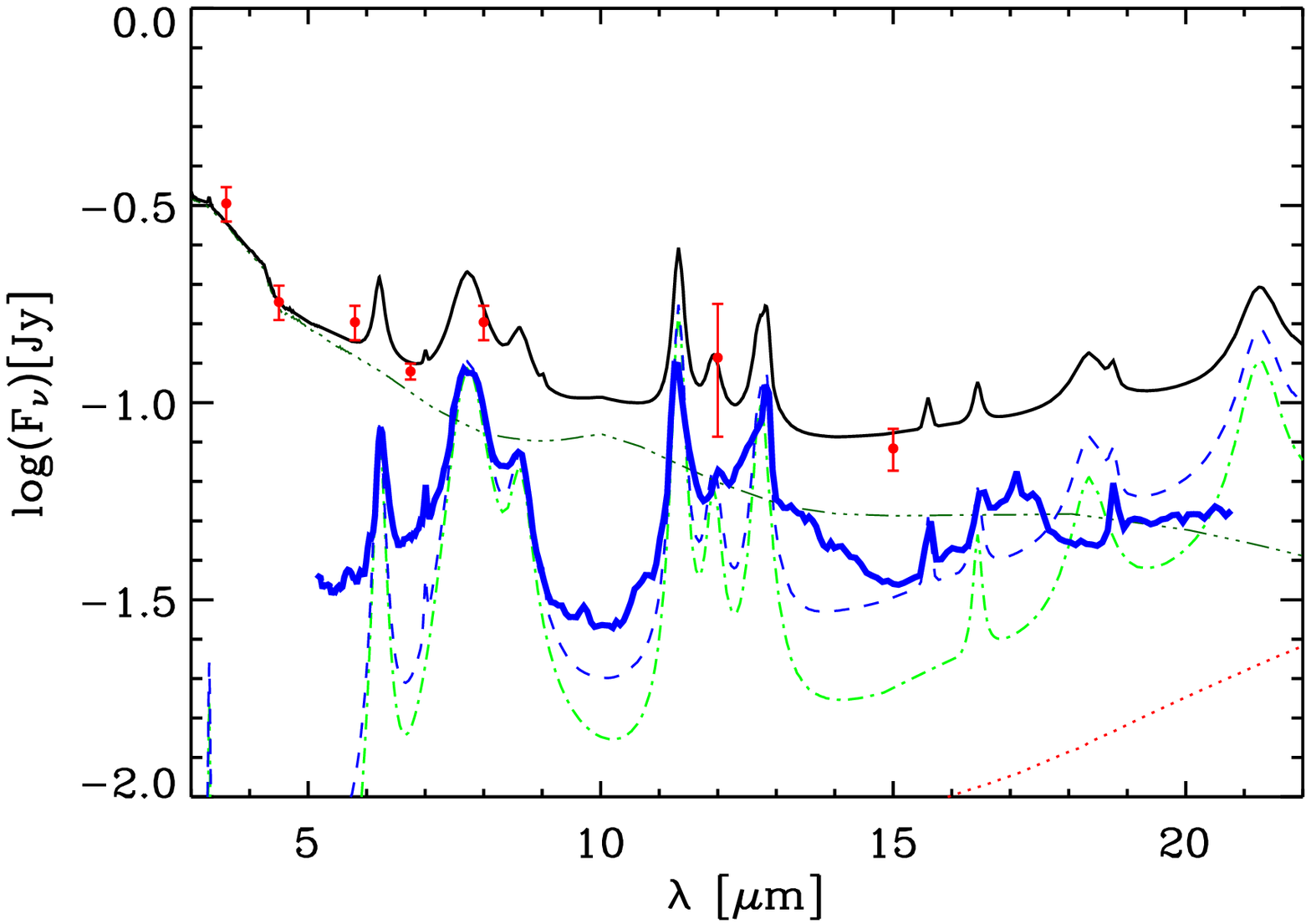}
\figcaption{Comparison between the total aperture SED of NGC~4435
and the best fitting model. See Figure \ref{fig:central_sed} for
the description of different lines.
\textit{Left}: Comparison from 0.1$\mu$m to 100 MHz.
\textit{Right}: 
Comparison for the MIR wavelengths.
The  thickest solid line represents the IRS \emph{Spitzer} spectrum.
\label{fig:total_sed}}
\end{figure*}

\subsection{Total aperture SED}

We have also compared our models with the 
total aperture SED of the galaxy. 
The total aperture SED includes
the U, B, V, J, H, and K bands from
GOLDMINE\footnote{Galaxy On Line Database Milano Network,
http://goldmine.mib.infn.it}; the LW2 (6.75~$\mu$m) and
LW3 (15~$\mu$m) ISO bands from \citet{bose03}; the UV (2000\AA) from Deharveng et al. (2002), 
all IRAS bands (with the exception of 25~$\mu$m), and the radio measurement described in section \ref{sec:sedglobal}.
These data are complemented by our IRAC flux measurements
and GALEX data for the full aperture. The data are reported in table 
\ref{tab:data2}.

The fit to the total aperture SED was done simply by changing the 
amount of the old stellar population and using the 
same starbust component that fits the central SED, 
as derived in the previous sections.

The resulting total mass of the old stellar
population is M$_{\rm{old}}^{\rm{Tot}}=4.66\times 10^{10}~M_\odot$. 
Moreover, the star forming component emits 8\%
of the total bolometric luminosity of the galaxy, 
while its stellar mass is only 0.3\% of the total 
stellar mass of the galaxy. It is worth noting that the starburst in 
N4438 (the late type companion) accounted for a smaller fraction 
(about 0.1\%) of its total stellar mass \citep{bose05}.

The comparison between the observed full aperture SED and the model is shown in the Figure
\ref{fig:total_sed}. 
The full aperture SED is well reproduced at all wavelengths. 
We stress that the old stellar component provides a significant 
contribution to GALEX, ISO
LW2, LW3 and IRAS 12$\mu$m broad band fluxes.
\begin{deluxetable}{lcccl}
\tablewidth{0pt}
\tablecaption{Total aperture broad band fluxes of NGC~4435\label{tab:data2}}
\tablehead{
\colhead{ $\lambda$ }& 
\colhead{Band /}& 
\colhead{Flux}&
\colhead{Uncert.}&
\colhead{Ref.}\\
\colhead{($\mu$m)}& 
\colhead{Instrument}&
\colhead{(Jy)}&
\colhead{(mJy)}&
\colhead{}
}
\startdata
0.2000&UV   &3.4E-4 &0.1 &1\\
0.3601&U    &0.0219 &4.38&2\\
0.4399&B    &0.0870 &13.05&2\\
0.5500&V    &0.166  &24.9&2\\
1.25  &J    &0.61   &122.0&2\\
1.65  &H    &0.70   &105.0&2\\
2.17  &K    &0.53   &106.0&2\\
3.6   &IRAC &0.32   &32.0&3\\
4.5   &IRAC &0.18   &18.0&3\\
5.8   &IRAC &0.16   &16.0&3\\
6.75  &LW2  &0.12   &4.61&4\\
8.0   &IRAC &0.16   &16.0&3\\
12    &IRAS &0.130  &29.2&5\\
15    &LW3  &0.0765 &8.73&4\enddata
\tablecomments{Complementary IRAS and radio fluxes were already reported in Table \ref{tab:data}.}
\tablerefs{(1) Deharveng et al. 2002; (2) Goldmine; (3) this work;
(4) Boselli et al. 2003; (5) NED.}
\end{deluxetable}


\subsection{X-rays}

As a final step in the SED analysis we briefly consider the X-ray
emission of NGC~4435, observed with {\em Chandra} by \citet{mach04}.
The X-ray spectrum was fitted with a thermal component in the soft (0.3-2
keV) band attributed to hot gas, and a non-thermal component responsible for the
2-10 keV band emission.  
The hard component, with a luminosity of
$L_{\rm{X}}=3.9\times 10^{32}$ W, is mostly concentrated within 3\arcsec\
from the galaxy center. This emission was attributed to  a
possible low luminosity AGN \citep[as suggested by][]{hofi97} plus recent
star formation activity.

The observed hard X-ray luminosity of NGC~4435 is about five times larger than 
that expected
from the FIR luminosity and the X-ray -- FIR correlation found for starbursts
galaxies \citep{rana03}.
Moreover, adopting the calibration by \citet{rana03}, the SFR 
implied by the observed hard X-ray luminosity turns out 
SFR$_{\rm{X-ray}}=0.78~M_\odot$~yr$^{-1}$, 
one order of magnitude 
higher than the current SFR derived from the fit to the global SED.

Though this seems to support the AGN nature of the hard emission
component we need to consider the peculiar star formation history of NGC~4435.
In fact, due to the much higher SFR at the beginning of the burst,
the average SFR is consistent with the X-Ray estimate.
For this reason we expect a significant contribution to the X-Ray luminosity
from intermediate mass stars accreting onto 
neutron stars and/or black holes, through Roche-lobe mass transfer.
To check this possibility we have computed the X-ray emission
expected from the star formation activity derived previously, 
with the synthesis model of  
\citet{silv03}. The resulting value is consistent
with the observed hard X-ray luminosity of NGC~4435.
Moreover, we notice that
the X-ray luminosity is much less dependent on the age of the 
stellar population (the turn-off mass of the stars)
than the FIR luminosity. This implies that, in post starburst galaxies,
the X-ray -- FIR luminosity ratio tends to be larger than that 
of pure starbursts which are dominated by massive stars.

Another indication of the strength of the AGN
comes from the comparison of the
hard X-ray luminosity and H$\alpha$ emission. 
Using the relation between $L_{\rm X}$-$L_{\rm H\alpha}$, 
derived for AGN \citep{hofe01}, the expected H$\alpha$
luminosity would be $L_{\rm H\alpha}^{\rm AGN}\sim 6.6\times 10^{31}$ W. 
This value is about 15\% of the observed $L_{\rm H\alpha}$
corrected for extinction via the Balmer decrement.
This contribution to the ionizing flux is higher 
than the limit derived from mid infrared 
nebular lines (Sect. \ref{sec:sf_agn}). However it must
be considered as a generous upper limit because
on one hand all the X-ray luminosity has been attributed to the AGN
and, on the other, it is known that the Balmer decrement 
provides only a lower limit to the attenuation \citep[e.g.][]{panu03}. 
In fact if we correct the observed $L_{\rm H\alpha}$ with the attenuation
derived from our best fit model we obtain an upper limit to the AGN contribution
$\sim$5\%, consistent with what is found from the 
[\ion{Ne}{iii}]/[\ion{Ne}{ii}] ratio.


\subsection{NGC~4435 and the phenomenology of the galaxy-galaxy interaction}

NGC~4435 is interacting with NGC~4438; in particular, simulations by
\citet{voll05} suggest that NGC~4435 passed through the disk of 
NGC~4438 about 100 Myr ago at a radial distance of
5-10 kpc. With such an impact parameter the interaction of the ISM of the
two galaxies is unavoidable. 

The age of the starburst $t_{\rm b}$ derived from our fit is 
larger than the (dynamically estimated) time since the
ISM-ISM collision (around 100 Myr), but it is 
consistent with the epoch of the onset of interaction.
This suggests that the star formation was
triggered by the interaction \citep[as often seen in galaxy-galaxy
simulations, e.g.][]{barn04} possibly producing a relevant
activity also before the ISM-ISM collision.
We cannot, however, draw strong conclusions 
on the relative importance of the tidal interaction and 
ISM-ISM collision, because 
it is difficult to derive the detailed 
shape of the star formation history during the burst.

Another phenomenon that may be triggered by
galaxy-galaxy interaction is AGN activity \citep[see e.g.][]{hopk06}.
However the various lines of evidence shown in this work  
indicate a lack of current AGN activity.

The age of the circumnuclear starburst is several times larger than the
typical duration of the AGN phase and we cannot rule out the possibility
that an AGN was once active in NGC~4435. However the surrounding gas 
is predominantly in a cold phase with only a 
fraction in a hot component \citep[$\simeq 3\times 10^7~M_\odot$,][]{mach04},
and it is supported by circular
motion \citep{cocc06}, indicative of a relaxed ISM. We conclude that the
feedback from the possible AGN was not strong enough to halt the star
formation process which is now fading as a result of the gas consumption.

Alternatively, the AGN activity has not been triggered yet (and
perhaps will not be triggered at all). 
\citet{cocc06} provide a 3$\sigma$ upper mass
limit of 7.5$\times 10^6~M_\odot$ for the possible central SMBH,
an order of magnitude lower than the value
obtained from the $M_{\rm BH}$-$\sigma$ and $M_{\rm BH}$-$L_{\rm bulge}$
relations. 
In this case NGC~4435 could be a prototype of the so
called ``laggard'' SMBH which are believed to undergo
slow gas fueling of the
galactic center and limited growth of the SMBH mass \citep{vitt05}.

The above picture sketched for the evolution of NGC~4435
would suggest that the star formation activity could inhibit the AGN
feeding both by consuming the gas and by SNe feedback.
Both cases point towards a relatively low efficiency of the AGN feedback
with respect to that of the SNe in intermediate/small mass galaxies
like NGC~4435 \citep[see e.g.][]{gran04}.


\section{Conclusions}
\label{sec:conclu}
We have performed a thorough spectrophotometric study of the 
early-type galaxy NGC~4435 using new \emph{Spitzer}
IRS spectra combined with IRAC and MIPS archival data and existing broad band
measurements from X-ray to radio wavelengths.

The IRS spectrum
is dominated by PAH features, nebular emission lines and molecular hydrogen
rotational lines that originate in the circumnuclear disk
that characterizes this galaxy.

The PAH features in the spectrum are very
similar to those observed in other star forming galaxies.
In particular, we report the detection of the PAH feature observed
at 17.1 $\mu$m \citep{smit04} which,
due to the lower contamination by the
H$_2$ S(1) rotational line, we show is actually
a blend of two distinct components.

The panchromatic SED of the central 5\arcsec\ region
was analysed with the spectrophotometric code GRASIL.

This SED is well reproduced from the UV to radio by a
model of a fading starburst superimposed on a simple stellar
population of 8 Gyr and solar metallicity.

The analysis of MIR nebular lines, combined with
that of the SED, allow a {\sl direct} estimate of the 
gas metallicity, which also results as solar.

Though a low level nuclear activity has been suspected in
NGC~4435 from optical \citep{hofi97,hosa02}
and X-ray \citep{mach04} observations, we collect the  
following evidence for the lack of AGN excitation:

{\em (i)} we have not detected high excitation
nebular emission lines in the MIR spectrum

{\em (ii)} the [\ion{Ne}{iii}]15.5/[\ion{Ne}{ii}]12.8 ratio
constrains the contribution of a possible AGN to the ionizing flux to be
less than 2\%.

{\em (iii)} the upper limit on the temperature derived from H$_2$ S(1) and S(2)
rotational lines is lower than expected for AGN excitation.

{\em (iv)} the X-ray emission is within the range expected from
X-ray binaries from the starburst

The age of the starburst, $186 \pm 37$ Myr, corresponds to the
epoch of the last encounter with NGC~4438 derived from dynamical simulations,
suggesting that this interaction has triggered the observed star formation 
event \citep{comb88}. However we cannot draw strong conclusions 
on the relative importance of the tidal interaction and 
ISM-ISM collision, because of the difficulty in deriving the detailed 
shape of the star formation history during the burst.

The combination of \emph{Spitzer} observations with the analysis of the
complete SED allows an accurate determination of the mass and epoch
of the rejuvenation episode.
The mass of the starburst ($\sim 1.40\times10^8~M_\odot$)
corresponds to about 1.4\% of the stellar mass
sampled by the central 5 arcsec aperture.
The young component makes the optical spectrum of NGC~4435 closely
similar to a typical interacting early-type galaxy with inverted
Ca{\sc\,ii}[H+K] lines \citep{long99,sarz05} that will
later turn into a typical {\sl cluster} E+A galaxy.
This study supports the notion
that early-type galaxies with 
relatively strong  hydrogen absorption features 
are due to recent small rejuvenation episodes, rather than
being the result of delayed galaxy formation \citep{bres96}.

Finally the total SED of NGC~4435 
is fitted by simply adding to the starburst a larger
fraction of the old stellar population. 
The starburst provides only 0.3\% of the total stellar mass of the galaxy
but 8\% of the bolometric luminosity.


\acknowledgments We are deeply indebted to A. Boselli and L.
Cortese for having provided us Galex UV data points. We thank E.
Corsini and L. Coccato for useful discussions and for the WFPC2
image of NGC~4435. We thank the anonymous referee for corrections and
helpful comments. 
We thank L. Paoletti, A. Petrella and D. Selvestrel 
for their work on the Galsynth interface.
A. B., G. L. G. and L. S. thank INAOE for warm
hospitality. O. V. acknowledges the support of the PICS MEXIQUE 2174
and the Mexican CONACYT projects  36547 E and 39714 F. This work is
based on observations made with the \emph{Spitzer} Space Telescope,
which is operated by the JPL, Caltech under a contract with NASA.


\end{document}